\documentclass[a4paper]{article}

\usepackage{INTERSPEECH2019}

\usepackage[utf8]{inputenc}
\usepackage{url}
\usepackage{color}

\title{Ultrasound-based Silent Speech Interface Built on a Continuous Vocoder}
\name{Tam\'as G\'abor Csap\'o$^{1,2}$, Mohammed Salah Al-Radhi$^{1}$, Géza Németh$^{1}$, \\
 G\'abor Gosztolya$^{3,4}$, Tam\'as Gr\'osz$^{4}$, L\'aszl\'o T\'oth$^{4}$, 
Alexandra Mark\'o$^{2,5}$}
\address{
  $^1$Department of Telecommunications and Media Informatics, \\
	Budapest University of Technology and Economics, Budapest, Hungary \\
	$^2$MTA-ELTE Lend\"ulet Lingual Articulation Research Group, Budapest, Hungary \\
	$^3$MTA-SZTE Research Group on Artificial Intelligence, Szeged, Hungary \\
	$^4$Institute of Informatics, University of Szeged, Hungary\\
  $^5$Department of Phonetics, E\"otv\"os Lor\'and University, Budapest, Hungary
 }
\email{\{csapot, malradhi, nemeth\}@tmit.bme.hu, \{ggabor, groszt, tothl\}@inf.u-szeged.hu, marko.alexandra@btk.elte.hu}

\begin{document}

\maketitle
\begin{abstract}
Recently it was shown that within the Silent Speech Interface (SSI) field, the prediction of F0 is possible from Ultrasound Tongue Images (UTI) as the articulatory input, using Deep Neural Networks for articulatory-to-acoustic mapping. Moreover, text-to-speech synthesizers were shown to produce higher quality speech when using a continuous pitch estimate, which takes non-zero pitch values even when voicing is not present.
Therefore, in this paper on UTI-based SSI, we use a simple continuous F0 tracker which does not apply a strict voiced / unvoiced decision. Continuous vocoder parameters (ContF0, Maximum Voiced Frequency and Mel-Generalized Cepstrum) are predicted using a convolutional neural network, with UTI as input.
The results demonstrate that during the articulatory-to-acoustic mapping experiments, the continuous F0 is predicted with lower error, and the continuous vocoder produces slightly more natural synthesized speech than the baseline vocoder using standard discontinuous F0.

\end{abstract}

\section{Introduction}

The articulatory movements are directly linked with the acoustic speech signal in the speech production process. Over the last decade, there has been significant interest in the articulatory-to-acoustic conversion research field, which is often referred to as “Silent Speech Interfaces” (SSI~\cite{Denby2010}). This has the main idea of recording the soundless articulatory movement, and automatically generating speech from the movement information, while the subject is not producing any sound. For this automatic conversion task, typically electromagnetic articulography (EMA)~\cite{Wang2012a,Kim2017a,Cao2018,Taguchi2018}, ultrasound tongue imaging (UTI)~\cite{Denby2004,Hueber2010,Hueber2011,Jaumard-Hakoun2016,Csapo2017c,Grosz2018,Toth2018,Tatulli2017}, permanent magnetic articulography (PMA)~\cite{Fagan2008,Gonzalez2017a}, surface electromyography (sEMG)~\cite{Nakamura2011,Janke2017,Wand2018}, Non-Audible Murmur (NAM)~\cite{Shah2018} or video of the lip movements~\cite{Hueber2010,Akbari2018} are used.

There are two distinct ways of SSI solutions, namely `direct synthesis' and `recognition-and-synthesis'~\cite{Schultz2017a}. In
the first case, the speech signal is generated without an intermediate step, directly from the articulatory data, typically
using vocoders~\cite{Cao2018,Taguchi2018,Denby2004,Hueber2011,Jaumard-Hakoun2016,Csapo2017c,Grosz2018,Gonzalez2017a,Janke2017}. In the second case, silent speech recognition (SSR) is applied on
the biosignal which extracts the content spoken by the person (i.e. the result of this step is text); this step is then followed by text-to-speech (TTS) synthesis~\cite{Wang2012a,Kim2017a,Hueber2010,Tatulli2017,Fagan2008,Wand2018}. 
In the SSR+TTS approach, any information related to speech prosody is totally lost, while several studies have shown that certain prosodic components may be estimated reasonably well from the articulatory recordings (e.g., pitch~\cite{Grosz2018,Nakamura2011,Liu2016,Zhao2017}).
Also, the smaller delay by the direct synthesis approach might enable conversational use.

Such an SSI system can be highly useful for the speaking impaired (e.g. after laryngectomy), and for scenarios where regular speech is not feasible, but information should be transmitted from the speaker (e.g. extremely noisy environments). Although there have been numerous research studies in this field in the last decade, the potential applications seem to be still far away from a practically working scenario.

\begin{figure*}
\centering
\includegraphics[trim=0.0cm 22.9cm 0.0cm 0.0cm, clip=true, width=0.85\textwidth]{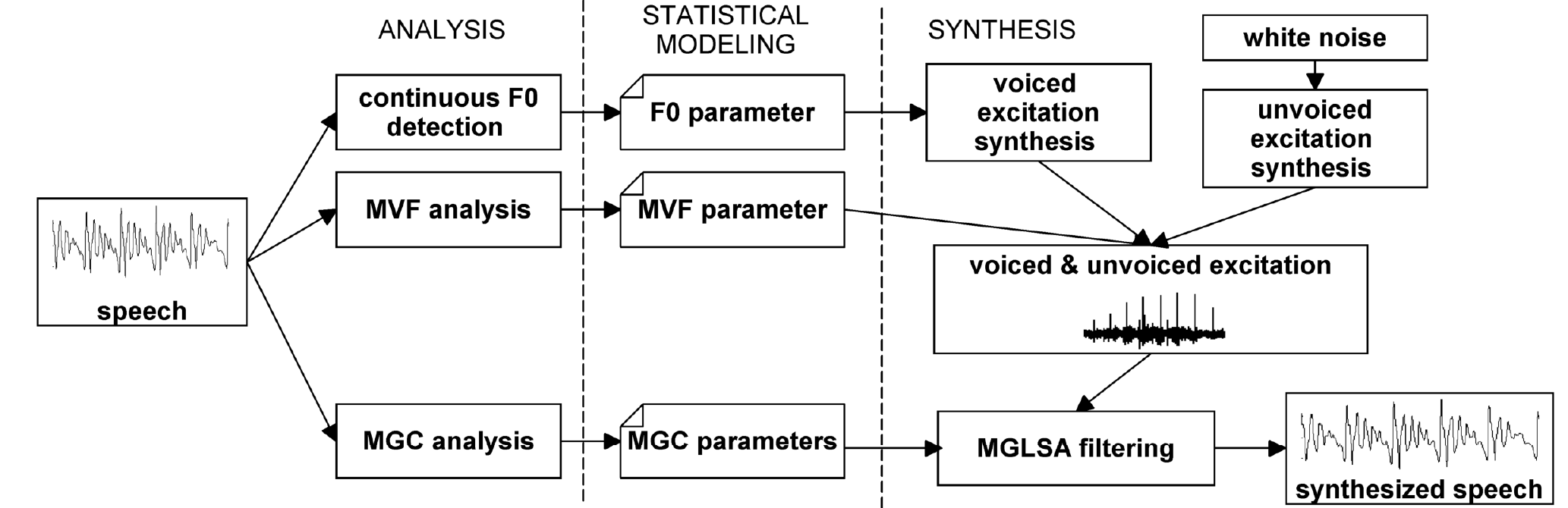}
\caption{General framework of the continuous vocoder.}
\label{fig:continuous_vocoder}
\end{figure*}

There have been only a few studies that attempted to predict the voicing feature and the F0 curve using articulatory data as input.
Nakamura et al.~used EMG data, and they divided the problem to two steps~\cite{Nakamura2011}. First, they used an SVM for voiced/unvoiced (V/UV) discrimination, and in the second step they applied a GMM for generating the F0 values. According to their results, EMG-to-F0 estimation achieved roughly 0.5 correlation, while the V/UV decision accuracy was 84\%~\cite{Nakamura2011}.
Hueber et al.~experimented with predicting the V/UV parameter along with the spectral features of a vocoder, using ultrasound and 
lip video as input articulatory data~\cite{Hueber2011}. They applied a feed-forward deep neural network (DNN) for the V/UV prediction and achieved 
82\% accuracy, which is very similar to the result of Nakamura et al. As the input data contained no direct measurements 
of vocal fold vibration, they explained this relatively high performance by indirect relationships (e.g. stable vocal tract configurations are likely to correspond to vowels and thus to voiced frames)~\cite{Hueber2011}.
Two recent studies experimented with EMA-to-F0 prediction. Liu et al.~compared DNN, RNN and LSTM neural networks 
for the prediction of the V/UV flag and voicing. They found that the strategy of cascaded prediction, 
that is, using the predicted spectral features as auxiliary input  
increases the accuracy of excitation feature prediction~\cite{Liu2016}.
Zhao et al. found that the velocity and acceleration of EMA movements are effective in articulatory-to-F0 prediction,
and that LSTMs perform better than DNNs in this task. Although their objective F0 prediction scores were promising, 
they did not evaluate their system in human listening tests~\cite{Zhao2017}.
In~\cite{Grosz2018}, we experimented with deep neural networks to perform articulatory-to-acoustic conversion from ultrasound images, with an emphasis on estimating the voicing feature and the F0 curve from the ultrasound input. We attained a correlation rate of 0.74 between the original and the predicted F0 curves, and an accuary of 87\% in V/UV prediction (when using five consecutive images as input). Also, subjective listening tests showed that the subjects could not differentiate between the sentences synthesized using the DNN-estimated or the original F0 curve, and ranked them as having the same quality. However, in several cases, the inaccurate estimation of the voicing feature caused audible artifacts.

\section{Continuous F0 modeling within vocoders}

In text-to-speech synthesis, the accurate modeling of fundamental frequency (also referred to as pitch or F0) is an important aspect. Its modeling in the statistical parametric speech synthesis framework with traditional approaches is complicated, because pitch values are continuous in voiced regions and discontinuous in unvoiced regions. For modeling discontinuous F0, MSD-HMM (Multi-Space Distribution - Hidden Markov model) was proposed~\cite{Tokuda2002} and it is generally accepted. However, because of the discontinuities at the boundary between voiced and unvoiced regions, the MSD-HMM is not optimal~\cite{Yu2011}. To solve this, among others, Garner and his colleagues proposed a continuous F0 model~\cite{Garner2013}, showing that continuous modeling can be more effective in achieving natural synthesized speech. 
Another excitation parameter is the Maximum Voiced Frequency (MVF), which was recently shown to result in a major improvement in the quality of synthesized speech~\cite{Drugman2014b}. During the synthesis of various sounds, the MVF parameter can be used as a boundary frequency to separate the voiced and unvoiced components. 

In our earlier work, we proposed a computationally feasible residual-based vocoder~\cite{Csapo2015d}, using a continuous F0 model~\cite{Garner2013}, and MVF~\cite{Drugman2014b}. In this method, the voiced excitation consisting of pitch synchronous PCA residual frames is low-pass filtered, while the unvoiced part is high-pass filtered according to the MVF contour as a cutoff frequency. The approach was especially successful for modeling speech sounds with mixed excitation. Next, we removed the post-processing step in the estimation of the MVF parameter and thus improved the modelling of unvoiced sounds within our continuous vocoder~\cite{Csapo2016}. Finally, we applied various time domain envelopes for advanced modeling of the noise excitation~\cite{Al-Radhi2017}.

In this paper, we use our continuous vocoder for ultrasound-based articulatory-to-acoustic mapping. Continuous vocoder parameters (ContF0, Maximum Voiced Frequency and Mel-Generalized Cepstrum) are predicted using a convolutional neural network, having UTI as input.

\section{Methods}
\label{sec:methods}

\subsection{Data acquisition}

Two Hungarian male and two female subjects with normal speaking abilities were recorded while reading sentences aloud (altogether 209 sentences each). The tongue movement was recorded in midsagittal orientation using the ``Micro'' ultrasound system of Articulate Instruments Ltd. at 81.67 fps. The speech signal was recorded with a Beyerdynamic TG H56c tan omnidirectional condenser microphone. The ultrasound data and the audio signals were synchronized using the tools provided by Articulate Instruments Ltd. More details about the recording set-up can be found in~\cite{Csapo2017c}. In our experiments, the scanline data of the ultrasound was used, after being resized to 64$\times$128 pixels using bicubic interpolation. The overall duration of the recordings was about 15 minutes, which was partitioned into training, validation and test sets in a 85-10-5 ratio.

\subsection{Baseline vocoder}

To create the speech synthesis targets, the speech recordings (digitized at 22\,050~Hz) were analyzed using an MGLSA vocoder~\cite{Imai1983} at a frame shift of 1 / (81.67 fps) = 270 samples, which resulted in F0, energy and 24-order spectral (MGC-LSP) features \cite{Tokuda1994}. The vocoder parameters served as the training targets of the DNN in our speech synthesis experiments. 

\subsection{Continuous vocoder}

The analysis and synthesis phases of the continuous vocoder are shown in Fig.~\ref{fig:continuous_vocoder}. It uses the same spectral representation (24-order MGC-LSP) as the baseline one. During analysis, ContF0 is calculated on the input waveforms using the simple continuous pitch tracker~\cite{Garner2013}. In regions of creaky voice and in case of unvoiced sounds or silences, this pitch tracker interpolates F0 based on a linear dynamic system and Kalman smoothing. After this step, MVF is calculated from the speech signal~\cite{Drugman2014b,Csapo2016}.

During the synthesis phase, voiced excitation is composed of residual excitation frames overlap-added pitch synchronously, depending on the continuous F0~\cite{Csapo2015d,Csapo2016,Al-Radhi2017}. After that, this voiced excitation is lowpass filtered frame by frame at the frequency given by the MVF parameter. In the frequency range higher than the actual value of MVF, white noise is used. Voiced and unvoiced excitation is added together. Finally, an MGLSA filter is used to synthesize speech from the excitation and the MGC parameter stream~\cite{Imai1983}.

\begin{figure*}
\centering
\includegraphics[trim=3.0cm 0.5cm 3.5cm 1.8cm, clip=true, width=1.0\textwidth]{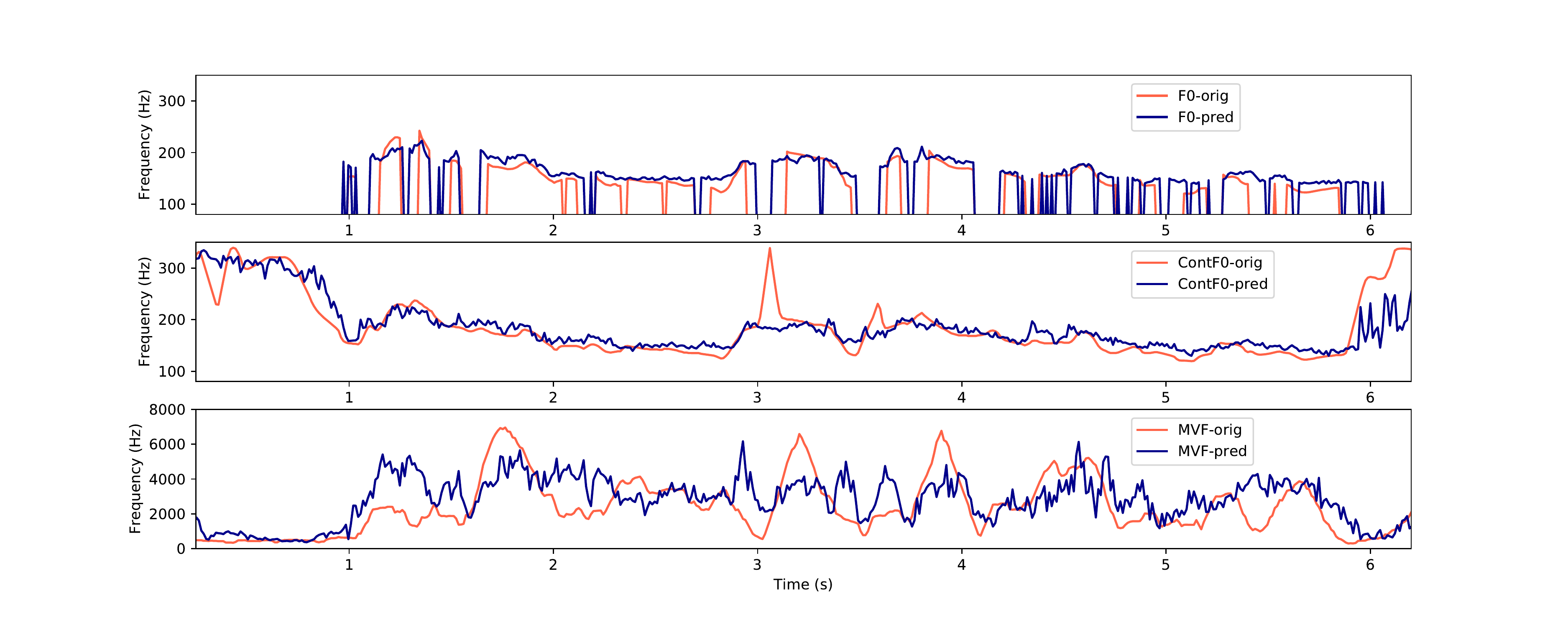}
\vspace{-4mm}
\caption{Demonstration samples from a female speaker. Top: F0 prediction using the baseline system. Middle: ContF0 prediction using the proposed system. Bottom: MVF prediction using the proposed system.}
\label{fig:proposed_sample}
\vspace{-4mm}
\end{figure*}

\subsection{DNN training with the baseline vocoder}

While in our earlier studies we were using simple fully-connected feed-forward networks~\cite{Csapo2017c,Grosz2018,Toth2018}, here we opted for convolutional neural networks (CNN), as typically they are more suitable for the processing of images. 
We trained speaker-specific CNN models using the training data of the four speakers (180~sentences from each of them).

Similarly to our earlier F0 prediction experiment~\cite{Grosz2018}, the baseline system consisted of two major machine learning components: one dedicated to make the voiced/unvoiced decision, while the role of the second was to estimate the actual F0 value for voiced frames. The first task, V/UV decision for each frame has a binary output, hence we handled it as a classification task. While working on the same input images, the second CNN seeks to learn the F0 curve. As the F0 estimate takes continuous values, we considered this task as a regression problem, and we trained this second network using only the voiced segments from the training data. 

For each speaker, altogether three neural networks were trained: one CNN for predicting the V/UV binary feature, one for predicting the log(F0), and one for predicting the 24-order MGC-LSP. All CNNs had one 64$\times$128 pixel ultrasound image as input, and had the same structure: two convolutional layers (kernel size: 3$\times$3, number of filters: 16 and 8, respectively), each followed by max-pooling, resulting in a dimension of 960. Finally, two dense layers were used with 1000 neurons each. In all layers, ReLU activation was used. The optimal parameters of the CNN architecture were calculated in an earlier hyperparameter optimization for the MGC-LSP target~\cite{Moliner2018}. Note that using several consecutive images as input, or applying recurrent architectures can lead to better results~\cite{Csapo2017c,Grosz2018,Moliner2018}, but here we did not apply these in order to test scenarios which are more suitable for real-time implementation.
The cost function applied for the log(F0) and MGC-LSP regression task was the mean-squared error (MSE), while for the V/UV classification we used cross-entropy. We trained the network using backpropagation, and applied early stopping to avoid over-fitting. The network was trained for 100 epochs and the training was stopped when the validation loss did not decrease within 10 epochs.

\subsection{DNN training with the continuous vocoder}

In the proposed system, three CNNs are used, with the same structure as the baseline system: two convolutional layers, each followed by max-pooling, and two dense layers with 1000 neurons each. The three different networks all have 64$\times$128 pixel images as input and are predicting the log(ContF0), log(MVF), and MGC-LSP features. The training procedures are the same as in the baseline, except that all three networks are performing regression, using the MSE cost function during training.

\section{Results and discussion}

\subsection{Demonstration sample}

A sample sentence (not appearing in the training data) was chosen for demonstrating how the baseline and proposed systems deal with the voicing and F0 prediction. Fig.~\ref{fig:proposed_sample} top shows the baseline system: first the V/UV decision was made by a CNN, and on the voiced frames, the F0 was predicted using another CNN. In general, the predicted F0 curve follows the original F0 line. However, the frequent V/UV changes result in a highly discontinuous F0 curve (e.g. between 4--5~s). Fig.~\ref{fig:proposed_sample} middle is an example for the ContF0 prediction: because the F0 was interpolated, there are F0 values even in the unvoiced sounds (e.g. at 1.6~s) or silences of the original sentence (e.g. between 0--1~s). In the proposed system, the MVF parameter (Fig.~\ref{fig:proposed_sample} bottom) is responsible for the voicing decision: if the MVF is low (below 1000~Hz), the resulting speech is more unvoiced, whereas if the MVF is high (above 1000~Hz), the synthesized speech has more voiced components. Therefore, the unnaturally high ContF0 values (e.g. between 0--1 and after 6~s in Fig~\ref{fig:proposed_sample}), which are the result of the interpolation, are compensated by the MVF parameter. As can be seen from the ContF0 and MVF prediction, the proposed system also has voicing errors, which typically result in false voicing. Overall, the main difference between the baseline and the proposed systems is the way how they handle the voicing feature of speech excitation.

\begin{table}[t]
\caption{Prediction performance of the CNNs on the test set.} \label{tab:objective}
\vspace{-2mm}
\centering
\renewcommand{\arraystretch}{1.2}
\begin{tabular}{l||c|c|c|c}
& \multicolumn{2}{c|}{{\bf Baseline}} & \multicolumn{2}{c}{{\bf Proposed}}  \\
\cline{2-5}
& {Acc. \%} & \multicolumn{3}{c}{{RMSE}}  \\
\cline{2-5}
{\bf Speaker~~~} &V/UV &~~~~F0~~~~& Cont.F0 &~~MVF~~\\
\hline\hline
Female \#1  & 81.41 & 71.75 & 30.24 & 1177.32 \\
Female \#2  & 77.28 & 83.30 & 31.24 & ~761.27 \\
Male \#1    & 74.84 & 47.05 & 28.18 & ~865.87 \\
Male \#2    & 81.84 & 59.25 & 32.57 & ~654.29 \\
\hline
Average     & 78.84 & 65.34 & 30.56 & ~864.69 \\
\end{tabular}
\vspace{-6mm}
\end{table}

\subsection{Objective evaluation}
\label{sec:objective}

After training the CNNs for each speaker and parameter individually, we synthesized sentences, and measured the accuracy of V/UV decision, and the Root Mean Square Error (RMSE) between the original and predicted F0 curves (for all segments) and MVF values. This objective evaluation was done on the test data (9 sentences from each speaker).
According to Table~\ref{tab:objective}, the V/UV decision accuracy was between 77--81\%, depending on the speaker, which is similar to earlier results in articulation-to-voicing prediction~\cite{Hueber2011,Nakamura2011,Grosz2018}. Note that in~\cite{Grosz2018} we achieved accuracies of 86--88\%, but that study used only one speaker (with roughly twice as large training data as here), while now we use four speakers. Comparing the standard F0 and ContF0 prediction, we can observe that the latter resulted in significantly lower RMSE (31~Hz as opposed to 65~Hz). The reason for this might be that the interpolated F0 is simpler to predict, as other studies also found earlier~\cite{Garner2013,Csapo2015d,Toth2016}. The RMSE of the Maximum Voiced Frequency prediction is in the range of 654--1177~Hz, indicating that for some speakers the MVF can be estimated with lower error, while for Female \#1 this task was more difficult. Comparing the UTI-to-MVF prediction results with text-to-MVF prediction (within HMM-based speech synthesis~\cite{Csapo2015d,Csapo2016}), the latter seems to be a simpler task. Predicting voicing related parameters from articulatory data can be done only through indirect relationships, as pointed out in Section~1. In HMM-TTS and DNN-TTS, typically a frame shift of 5~ms is used, whereas here we are tied to the frame rate of the ultrasound (82 FPS, resulting in 12~ms frame rate).

\subsection{Subjective listening test}

In order to determine which proposed system is closer to natural speech, we conducted an online MUSHRA (MUlti-Stimulus test with
Hidden Reference and Anchor) listening test~\cite{mushra}.
Our aim was to compare the natural sentences with the synthesized sentences of the baseline, the proposed approaches and a benchmark system (the latter having constant F0 across the utterance). We included cases where the excitation parameters were predicted and original MGC-LSP was used during synthesis, and also cases where all vocoder parameters were predicted from the ultrasound images. In the test, the listeners had to rate the naturalness of each stimulus in a randomized order relative to the reference (which was the natural sentence), from 0 (very unnatural) to 100 (very natural). We chose three sentences from the test set of each speaker (altogether 12 sentences). The samples can be found at \url{http://smartlab.tmit.bme.hu/interspeech2019_ssi_f0}.

\begin{figure}
\centering
\includegraphics[trim=0.2cm 0.3cm 0.3cm 0.2cm, clip=true, width=\columnwidth]{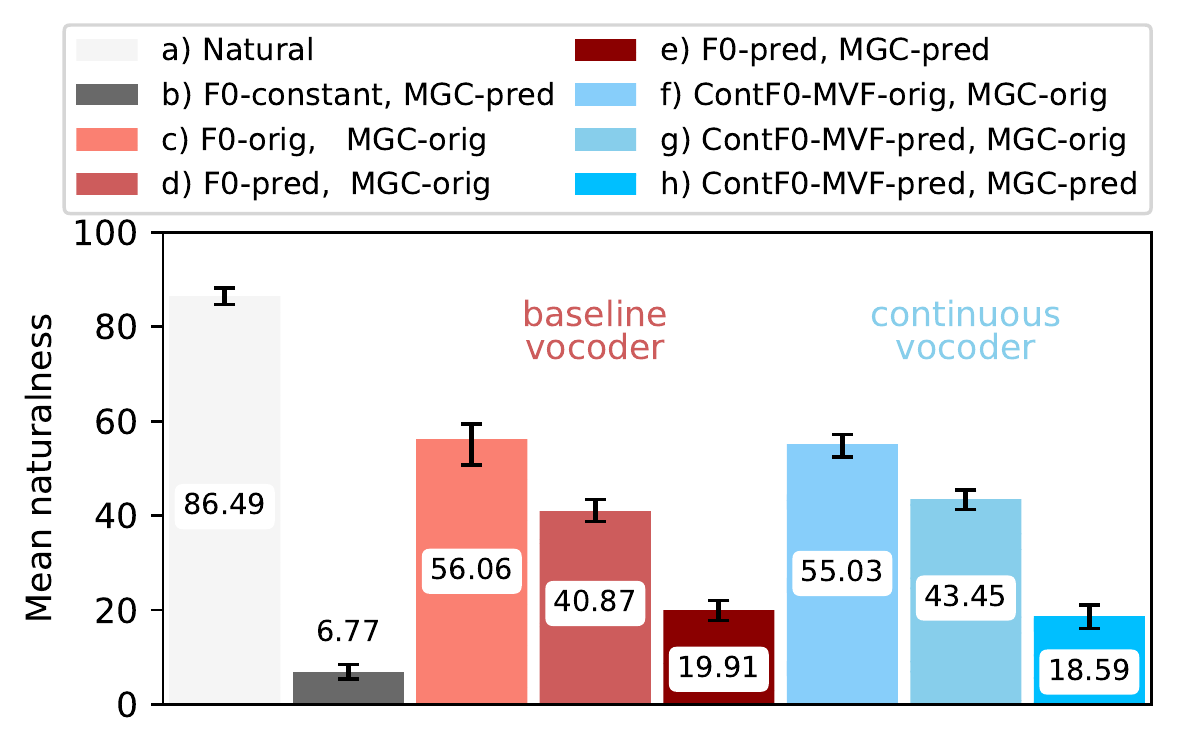}
\vspace{-5mm}
\caption{Results of the subjective evaluation for the naturalness question. 'orig'=original, 'pred'=predicted by the CNN.}
\label{fig:results_subjective}
\vspace{-6mm}
\end{figure}

Each sentence was rated by 23 native Hungarian speakers (9~females, 14 males; 19--47 years old). On average, the test took 15 minutes to complete. Fig.~\ref{fig:results_subjective} shows the average naturalness scores for these tested approaches. The benchmark version (with constant F0) achieved the lowest scores, while the natural sentences were rated the highest, as expected. The variants in which the sentences were analyzed with the baseline and proposed vocoders (without CNN training) were rated equal ('c' and 'f'). In cases where the excitation parameters (V/UV, F0, ContF0, or MVF) were predicted by the CNNs but the original MGC-LSP features were used during synthesis, the preference towards the proposed methods is visible (in absolute terms) : the variant 'g' achieved 43.45\% compared to the baseline system 'd' of 40.87\%. Finally, when all vocoder parameters were predicted, the two systems were rated roughly equal ('e' and 'h'). To check the statistical significance of the differences, we conducted Mann-Whitney-Wilcoxon ranksum tests with a 95\% confidence level, but none of the 'c'--'f' / 'd'--'g' / 'e'--'h' pairs differed significantly. As a summary of the listening test, a slight (but not significant) preference towards the proposed vocoder could be observed when the CNN-predicted ContF0 and MVF values were used with the original spectral features.

\section{Conclusions}

Here, we described our experiments to perform continuous F0 and Maximum Voiced Frequency estimation in articulatory-to-acoustic mapping. We used separate speaker-dependent convolutional neural networks to predict the ContF0, MVF and MGC-LSP parameters from Ultrasound Tongue Image input.
The results of the objective evaluation demonstrated that during the articulatory-to-acoustic mapping experiments, the continuous F0 is predicted with significantly lower RMS error than the discontinuous F0 (31~Hz vs.\ 65~Hz). According to the subjective test, the continuous vocoder produces slightly (but not significantly) more natural synthesized speech than the baseline.

The advantage of this continuous vocoder is that it is relatively simple: it has only two 1-dimensional parameters for modeling excitation (ContF0 and MVF) and the synthesis part is computationally feasible, therefore speech generation can be performed in real-time. In our earlier UTI-to-F0 experiment~\cite{Grosz2018}, two types of DNNs were necessary: one for the V/UV classification, and another one for the F0 regression. With this new vocoder, a single CNN could be used to train all continuous parameters, which can result lower RMS errors, as shown in~\cite{Grosz2018} for the case of F0 and MGC-LSP.

\section{Acknowledgements}

The authors were partially funded by the NRDIO of Hungary (FK 124584, PD 127915 grants) and by the EFOP-3.6.1-16-2016-00008 grant. G.~Gosztolya and L.~Tóth were funded by the J\'anos Bolyai Scholarship of the Hungarian Academy of Sciences, by the Hungarian Ministry of Innovation and Technology New National Excellence Program ÚNKP-19-4, and by the project TUDFO/47138-1/2019-ITM.

\clearpage

\bibliographystyle{IEEEtran}

\bibliography{ref_collection_csapot_nourl}

\end{document}